\documentclass[11pt,twoside]{article}
\usepackage{asp2014}

\resetcounters

\begin{document}
\title{The population of white dwarf binaries with hot subdwarf companions}
\author{S. Geier$^1$, T. Kupfer$^2$, U. Heber$^3$, B. N. Barlow$^4$, P. F. L. Maxted$^5$, C. Heuser$^3$, V. Schaffenroth$^{3,6}$, E. Ziegerer$^3$, R. H. \O stensen$^{7}$, B. T. G\"ansicke$^8$
\affil{$^1$European Southern Observatory, Karl-Schwarzschild-Str. 2, 85748 Garching, Germany}
\affil{$^2$Department of Astrophysics/IMAPP, Radboud University Nijmegen, P.O. Box 9010, 6500 GL Nijmegen, The Netherlands}
\affil{$^3$Dr. Karl Remeis-Observatory \& ECAP, Astronomical Institute,
Friedrich-Alexander University Erlangen-Nuremberg, Sternwartstr. 7, D~96049 Bamberg, Germany}
\affil{$^4$Department of Physics, High Point University, 833 Montlieu Avenue, High Point, NC 27262, USA}
\affil{$^5$Astrophysics Group, Keele University, Staffordshire, ST5 5BG, UK}
\affil{$^6$Institute for Astro- and Particle Physics, University of Innsbruck, Technikerstr. 25/8, 6020 
Innsbruck, Austria}
\affil{$^7$Institute of Astronomy, K.U.Leuven, Celestijnenlaan 200D, B-3001 Heverlee, Belgium}
\affil{$^8$Department of Physics, University of Warwick, Coventry CV4 7AL, UK}}

\begin{abstract}
Hot subdwarfs (sdBs) are core helium-burning stars, which lost almost their entire hydrogen envelope in the red-giant phase. Since a high fraction of those stars are in close binary systems, common envelope ejection is an important formation channel. We identified a total population of 51 close sdB+WD binaries based on time-resolved spectroscopy and multi-band photometry, derive the WD mass distribution and constrain the future evolution of these systems. Most WDs in those binaries have masses significantly below the average mass of single WDs and a high fraction of them might therefore have helium cores. We found 12 systems that will merge in less than a Hubble time and evolve to become either massive C/O WDs, AM\,CVn systems, RCrB stars or even explode as supernovae type Ia. 
\end{abstract}

\section{Introduction}

After finishing core hydrogen-burning the progenitors of hot subdwarf stars (sdBs) leave the main sequence and evolve to red giants before igniting helium and settling down on the extreme horizontal branch. Unlike normal stars, the sdB progenitors must have experienced a phase of extensive mass loss on the red giant branch. What causes this extensive mass loss remains an open question. About half of the sdB stars reside in close binaries with periods ranging from one hour to a few days \citep{maxted01,napiwotzki04}. Because the components' separation in these systems is much less than the size of the subdwarf progenitor in its red-giant phase, these systems must have experienced a common-envelope and spiral-in phase \citep{han02,han03}. 

Most of the close companions to sdB stars that could be classified so far are low-mass main sequence stars, brown dwarfs or white dwarfs (WDs). More massive compact companions like neutron stars or black holes have been predicted by theory and some candidates have been found \citep{geier10a,geier10b}. Subdwarf binaries with massive WD companions turned out to be candidates for supernova type Ia (SN~Ia) progenitors because these systems lose angular momentum due to the emission of gravitational waves and shrink. Stable mass transfer or the subsequent merger of the system may cause the WD to explode as a SN~Ia \citep{maxted00,geier07,geier13}. 

Here we identify and investigate the known population of close sdB+WD binaries, derive the WD mass distribution and constrain the future evolution of these systems (see Kupfer et al. in prep. for details). 

\section{Selection of the sample}

Several studies have been performed to measure radial velocity curves and derive the orbital parameters of short-period sdB binaries \citep[e.g.][]{morales03, copperwheat11}. In the MUCHFUSS (Massive Unseen Companions to Hot Faint Underluminous Stars from SDSS) project, we derived orbital and atmospheric parameters of 30 sdB binaries \citep[e.g.][]{geier11a,geier11b,geier14}.
In total, 141 short-period sdB binaries have measured radial velocity curves. In most cases it is hard to constrain the nature of the companion as most sdB binaries are single-lined systems and
thus only the mass function $$f_{\rm m} = \frac{M_{\rm WD}^3\sin^3i}{(M_{\rm WD} + M_{\rm sdB})^2} = \frac{P K^3}{2 \pi G}$$ can be calculated. Although the RV semi-amplitude $K$ and the orbital period $P$
can be measured from the RV curve, $M_{\rm sdB}$, $M_{\rm WD}$ and the inclination angle $i$ remain free parameters. Assuming the canonical value of $0.47\,M_{\rm \odot}$ for $M_{\rm sdB}$ only a lower limit can be derived for $M_{\rm WD}$.

The best way to identify compact companions in those binaries is to exclude the presence of cool, late-type objects like M-dwarfs or brown dwarfs. We used two different techniques to select our sample of close sdB+WD binaries. 

The hemisphere of a cool low-mass companion facing the sdB is heated up by the significantly hotter sdB star. This causes a sinusoidal variation in the light curve. More/less flux is emitted if the irradiated hemisphere of the cool companion is faced towards/away from the observer. If this so-called reflection effect is detected, a compact companion can be excluded. However, if the light curve of the short period system shows no variation, a compact object like a WD is most likely to be the companion. In some cases, small ellipsoidal variations, Doppler boosting or eclipses reveal the presence of a WD companion directly. We found 22 binaries fulfilling those criteria.

A cool companion of spectral type $\sim\,M$1$-M$2 and earlier is detectable from an infrared excess even if the spectra in the optical range are not contaminated with spectral lines from the cool companion \citep{stark03, reed04}. Hence, the presence of a cool companion can be inferred by its infrared excess. We inspected each system with 2MASS/UKIDSS ($J$ and $H$) and $V$ band color information for an infrared excess to put tighter constraints on the nature of the companions. If a system does not show an excess in the infrared a cool companion can be excluded when the minimum companion mass derived from the RV-curve is higher than the mass of a stellar companion which would cause such an excess. In this way we identified another 29 sdB+WD binaries.

\begin{figure*}[t]
\begin{center}
 \includegraphics[width=13cm]{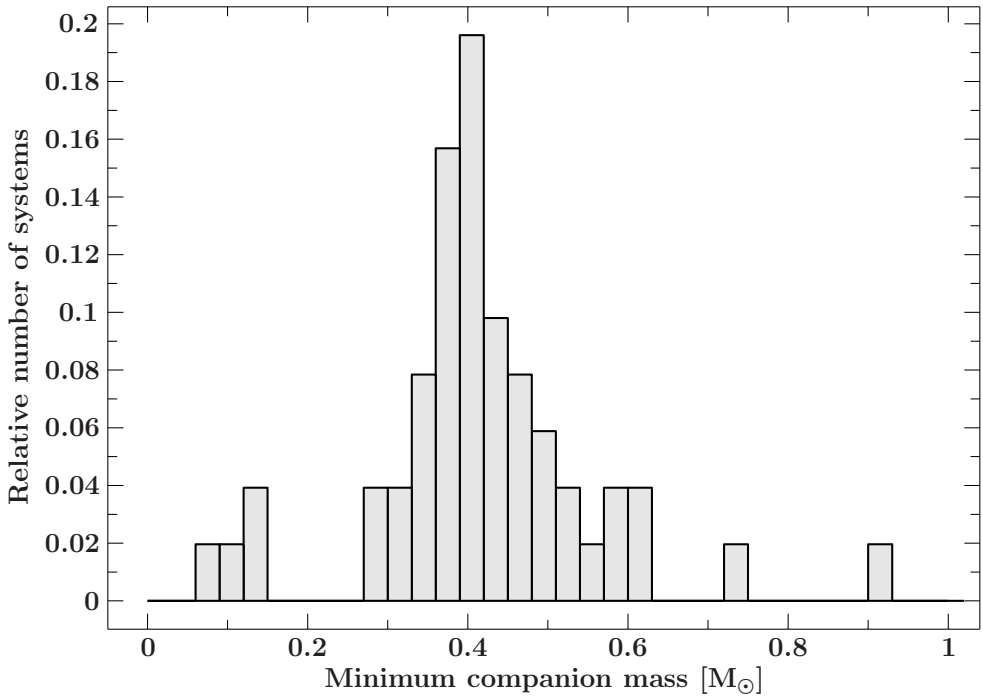}
 \caption{Histogram of minimum WD companion masses.}
 \label{fig:comp_mass}
\end{center}
\end{figure*}

\begin{figure*}[t]
\begin{center}
 \includegraphics[width=13cm]{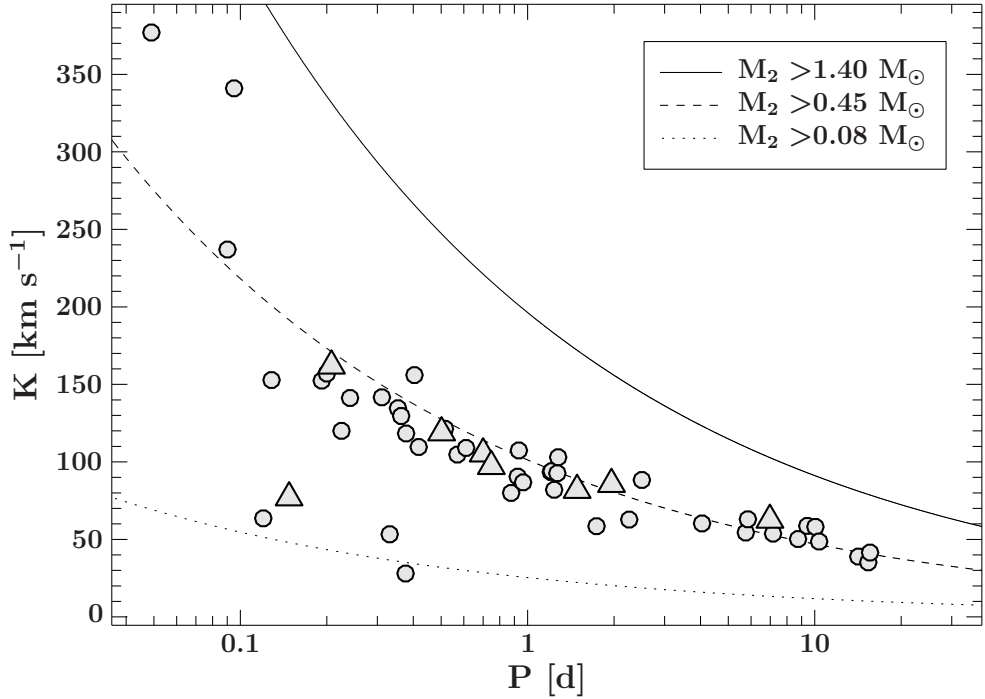}
 \caption{The RV semi-amplitudes of all known 51 short-period sdB+WD binaries with spectroscopic solutions plotted against their orbital periods. The binaries from the MUCHFUSS project are marked with triangles, binaries taken from the literature with circles. The lines mark the regions to the right where the minimum companion masses derived from the binary mass function (assuming $0.47\,M_{\rm \odot}$ for the sdBs) exceed certain values.}
 \label{fig:much_mass}
\end{center}
\end{figure*}

\section{WD mass distribution}

The distribution of the WD minimum masses of the 51 sdB+WD binaries is displayed in Fig.~\ref{fig:comp_mass}. Due to projection effects and selection biases the detection of high inclination systems should be favoured, which means that the derived limits should be on average close to the real WD masses. The majority of this population has an average mass of around $0.4\,M_{\rm \odot}$. This mass is significantly lower than the average mass of single WDs ($\sim0.6\,M_{\rm \odot}$), which leads to the conclusion that the WDs need to lose a significant amount of mass during the evolution either during the first phase of mass transfer when the WD is formed or during the common envelope phase when the sdB is formed. White dwarf masses of $\sim0.4\,M_{\rm \odot}$ are in the border region between WDs with helium cores and C/O cores and a significant fraction of the WD companions might be helium core WDs. Four WDs might even be of extremely low mass (ELM-WDs, $M<0.3\,M_{\rm \odot}$). The third group are the high mass WD companions ($>0.7\,M_{\rm \odot}$).

\section{Future evolution}

The orbits of the sdB+WD binaries are predicted to shrink due to the loss of angular momentum carried away by gravitational wave radiation until the two stellar remnants come into contact and start to transfer mass. Taking this effect into account, one can constrain the future evolution of these systems. Merger timescales for all the close binaries were calculated following \citet{pac67}. We identified only 12 sdB+WD systems which will merge within a Hubble time. The sdBs will evolve to become C/O WDs before contact. Depending on the mass ratio the systems either merge ($q>2/3$, 8 binaries) or form an AM\,CVn type binary ($q<2/3$, 2 binaries). For a helium white dwarf companion the merger might form an RCrB star, whereas a C/O-WD companion forms a massive single C/O-WD. Furthermore, two of the sdB+WD binaries \citep{maxted00,geier07,geier13} are already known as progenitors of SN\,Ia \citep{webbink84}.

\bibliography{geier}

\end{document}